\documentclass[a4paper,11pt]{article}
\pdfoutput=1
\usepackage{jcappub} 
\usepackage[T1]{fontenc}
\usepackage[titletoc,title]{appendix}
\newcommand{\beq}{\begin{equation}}
\newcommand{\eeq}{\end{equation}}
\usepackage{graphicx}
\graphicspath{{./Figures/}}
\usepackage{amsmath}
\usepackage{relsize}
\usepackage{MnSymbol}
\usepackage{mathrsfs}
\usepackage{subcaption}
\usepackage{mwe}
\usepackage{gensymb}

\newcommand{\sigmax}{$\sigma_{\rm max}$}

\begin{document}

\title{\boldmath Energy loss during Dark Matter propagation in an overburden}
\author{M. Shafi Mahdawi and Glennys R. Farrar}
%\author{}
\affiliation{Center for Cosmology and Particle Physics, Department of Physics, New York University, \\4 Washington Place, New York, NY 10003, USA}
\emailAdd{shafi.mahdawi@nyu.edu}
\emailAdd{gf25@nyu.edu}

\abstract{As experimental constraints on DM interactions become ever more sensitive and push into new regimes of DM mass, it becomes more and more challenging to accurately model the process by which Dark Matter particles lose energy through scattering in the Earth's surface or other overburdens.  We show that a commonly-used approximation due to Starkman, Gould, Esmailzadeh and Dimopoulos (SGED) can fail badly in computing the attenuation, even while being useful for an order-of-magnitude estimate of the maximum cross section reach.  We introduce a method of importance sampling which makes Monte-Carlo simulation of energy loss feasible, in spite of factor-$10^7$ or greater attenuation. We demonstrate the validity of our new method and expose multiple problems with the SGED approximation; this reveals interesting features of the energy loss process. We spot-check the recent Emken, Kouvaris and Shoemaker ``$5\,\Delta v$'' prescription to place limits on cross sections based on a limited-statistics analysis, and find that an accurate simulation yields a factor of $\rm{4.4\times10^6}$ and $\rm{2.4\times10^4}$ larger number of events, for 50 MeV and 1 GeV DM mass respectively, than if the EKS $5\,\Delta v$ prescription were valid.}
\maketitle
\flushbottom
\section{Introduction}	
If dark matter particles have a cross section large enough that they scatter in some overburden prior to reaching a DM detector, they can lose such a large fraction of their primary energy en route to the detector, that they cannot produce an energy deposit above threshold~\cite{Starkman}. %Furthermore, a significant fraction of the DM particles, typically $\approx 80$\% for well-matched masses, are simply reflected prior to reaching the detector.  
Thus a direct detection experiment at a given detector-depth is insensitive to Dark Matter having a DM-nucleus cross section larger than some value, which depends on DM mass, due to the reduction in flux of DM particles that are capable of triggering the detector.  We will call the upper-limit on the cross section sensitivity of that experiment, its \emph{maximum cross section reach} or $\sigma_{\rm max}$; $\sigma_{\rm max}$ is a function of DM mass $m$.
\par
A detailed Monte Carlo simulation is required to accurately determine the $\sigma_{\rm max}$ corresponding to a given experimental measurement.  However with attenuation factors of order $10^7$ or more, doing the required computations with a brute-force approach is often computationally infeasible. The attenuation factor is the number of DM particles which would have been detected during the experiment in the absence of an overburden, divided by the number of particles observed passing the threshold condition in the actual experiment. Thus, as the sensitivity of direct detection experiments increases, the attenuation factors encountered in interpreting the experimental data will become ever larger until DM is detected.

In this paper, we present a new approach to doing such propagation simulations which reduces the computation time by a factor $\mathcal{O}(10^3)$ or more, making the required computations feasible. To motivate the method, we must examine in some detail the properties of the successful DM particle trajectories. This gives useful physical insight and at the same time helps one to understand why the approaches used to date are inaccurate. We developed our method in order to re-examine the window for moderately-interacting DM ($\sigma_p \sim \mu$b where $\sigma_p$ is DM-nucleon cross section) in the $\sim$GeV mass range~\cite{bounds}.
\par	
In order to estimate the flux of capable\footnote{We designate as \emph{capable}, those particles reaching the detector with enough energy to potentially be successful if they interact, and as \emph{successful}, those particles causing an energy deposit above threshold in the detector.  Depending on the question, one or the other class may be most relevant.} particles, so as to extract cross section limits in the regime of strongly interacting dark matter,  Starkman, Gould, Esmailzadeh, and Dimopoulos~\cite{Starkman} (SGED) introduced an analytic approximation to the energy loss. The essence of the approximation is the assumption that all particles suffer a number of collisions equal to the number of interaction lengths to the detector, and that the fractional energy loss in each collision is the mean value averaged over isotropic CM scatterings (or whatever the fundamental differential cross section may be). Additionally, SGED assumed a single, linear path equal to the vertical depth of the detector, and took all DM particles to have an initial energy equal to the maximum over the distribution.  A more accurate treatment can easily be obtained by improving on the latter two simplifications.
\par	
However a more fundamental problem in using the SGED approximation to actually determine the maximum cross section reach of a particular experiment, lies in its underlying assumption of continuous, infinitesimal energy loss, and use of the mean interaction length and the mean energy loss per interaction, en route to the detector. %These mean-value approximations are valid only when the number of interactions is so large that  deviations from the central limit theorem have negligible effect.  This is the case when the maximum energy loss per collision is very small compared to the energy loss needed to make the particle undetectable.  
As we show below, these approximations are either not valid or improperly implemented in two recent works which aimed to extract the limit of sensitivity of the DAMIC --- Dark Matter In CCDs ---  experiment~\cite{Barreto2012264}.  The first, by Kouvaris and Shoemaker~\cite{Kouvaris2014} (KS), was for DM masses above $\sim 1$ GeV, and the second, by Emken, Kouvaris and Shoemaker (EKS)~\cite{Emken}, was for masses $\sim50$ MeV to $\sim1$ GeV.   
\par
We show in section~\ref{SGED} that, when the energy loss per collision is large -- of order 10\% for GeV-mass DM particles in the crust of the Earth -- the actual trajectories of \emph{capable} and \emph{successful} DM particles are strongly biased away from the SGED, mean-behavior description.  Trajectories of both successful and capable particles have many fewer scatterings than expected from the mean interaction length, and the CM scattering angles are skewed toward forward scattering, resulting in lower than average energy loss and shorter total path length compared to the entire ensemble of particles reaching the detector. For a given cross section, many more particles reach the detector with sufficient energy, than expected from the SGED approximation using mean energy loss and scattering length.  
\par
A detailed simulation is even needed when the energy loss per collision is very small, so that many collisions can be tolerated before the particle becomes incapable of an above-threshold energy deposit in the detector. In this case the continuous energy loss approximation is valid, but calculating it correctly requires a detailed simulation. EKS performed a simulation of the trajectories, but used a shortcut to extrapolate to the large-energy-loss tail resulting in a substantially wrong value of attenuation, as discussed in section~\ref{EKS}. Even in this low energy-loss case, a more efficient computational method is needed.

%If energy loss is important at all, then an improvement to the SGED/KS/EKS approximation is needed in order to extract accurate cross sections or cross section limits from direct detection experiments. The most reliable approach is to simulate DM trajectories to find the flux of capable particles at the detector depth, as a function of DM mass, cross-section and the detector energy threshold.  However in some cases of interest, the flux of capable particles is attenuated by almost a factor $10^9$, making the computational cost prohibitive.  Therefore we developed an importance sampling technique which reduces the computational cost by a factor $\mathcal{O}(10^3)$ or more. 

\par
After the intuition-building and motivational preambles of sections~\ref{SGED} and~\ref{EKS}, in section~\ref{IS} we explain the importance sampling %technique and discuss the specific importance sampling
scheme we developed.  In section~\ref{valid} we demonstrate its validity and accuracy.

\section{The SGED approximation and its limitations}\label{SGED}
Starkman et al.~\cite{Starkman} (SGED) proposed a continuous energy-loss and small-deflection angle approximation for strongly interacting DM particles. This method was applied by Kouvaris and Shoemaker~\cite{Kouvaris2014}, KS hereafter, to analyze results from DAMIC~\cite{Barreto2012264}. In this section, we describe the SGED/KS approximation and show that it fails by orders of magnitude to give the correct attenuation factor, especially when the recoil energy in a given DM interaction can be comparable to DM particle energy itself, i.e., when the mass of the DM particle and target are relatively well matched. For current detector thresholds, the minimum number of collisions in the overburden required to make the particle undetectable, is small enough that fluctuations strongly influence the final energy.
\par
Simple kinematics dictates that in a non-relativistic, elastic collision with a particle at rest, the energy loss of the scattered projectile is just its initial energy times a function of DM mass $m$, target nucleus mass $m_A$ (reduced mass $\mu_A$) and center-of-mass scattering angle $\xi_{CM}$:
\beq
\begin{split}
	\label{DeltE}
	\Delta \, E &= \frac{4\,\mu^2_{A}}{m\,m_A} \,\left(\frac{1-\cos\,\xi_{CM}}{2} \right)  \, E\\
	&\equiv f \, E .  
\end{split}
\eeq
%(See Appendix for more details.)
That is, the \emph{fractional} energy loss is independent of the energy.  Letting $\langle f\rangle$ denote the mean fractional energy loss (i.e., averaged over scattering angles), we expect
\beq
\label{EfEi}
 \langle E_f\rangle = (1-\langle f\rangle)^N E_i
\eeq
where $N$ is the number of scatterings and $E_{i,f}$ are the initial and final DM particle energies, respectively.  For eq.~\eqref{DeltE} to be valid, $N$ must be large enough that the variance in $f$ can be ignored.
\par
The SGED approximation goes further, treating the energy-loss as continuous, so that
\beq
\begin{split}\label{edEdz_t}
	\frac{dE}{dz}&=-\,\frac{\langle f\rangle \, E}{\lambda}\;,
\end{split}\eeq
where $\lambda\equiv (n\,\sigma)^{-1}$ is the interaction length of DM particles in the shielding material.  If applicable, this continuous treatment has the practical advantage of giving a simple closed-form expression for the final energy after traveling a distance $L$:
\beq
\label{SGEDESimple}
E(L) = E(0) \, \exp \left(-\, \langle f\rangle\,\frac{L}{\lambda} \right)~.
\eeq
If the maximum fractional energy loss is very small, then this continuous approximation can be valid.  
Whether using~\eqref{EfEi} or \eqref{SGEDESimple}, one must know $\langle f\rangle$ and $\langle N\rangle$, or in the SGED formulation, $L/\lambda \equiv \langle N\rangle$. As we show below, determining the pertinent values of these quantities is non-trivial, but we put that aside for a moment.  
\par
Assuming  $\langle f\rangle$ and $L$ are known, SGED estimates the maximum cross section reach, as follows. Let $E_{min}$ be the minimum energy of a DM particle to trigger the detector in question, given the threshold recoil energy of the detector, and let $E_{max}$ be the maximum energy a DM particle can have at the surface of the Earth, given the escape velocity of the Galaxy. Then, if eq.~\eqref{SGEDESimple} is applicable, no DM particles will have sufficient energy to trigger the detector unless the cross section is less than 
\beq
\label{bound}
\sigma_{A}^{max} = \left( n\, L\, \langle f\rangle \,\right)^{-1}\,\ln\left(\frac{E_{max}}{E_{min}}\right)\;.
\eeq
The relation between $\sigma_A$ and $\sigma_p$ is given in eq.~\eqref{e_sigmaAF2}. If this limit is saturated or exceeded, the experiment should see no events giving energy deposits above threshold.  
\par
One by one, let us consider the various deficiencies in the above approach, saving the most crucial ones to the last (numbers 4 and 5).  
\begin{enumerate}
\item
The crude SGED/KS method calculates the DM-proton cross section for which only the most energetic DM particles before passing through the shielding material are capable of triggering the detector.  This is easily improved, to take into account the actual DM spectrum and the observed number of events, while still using eq.~\eqref{SGEDESimple} for the energy loss. Details are given the Appendix where it is reported that the crude SGED\textbf{/}KS max-min method overestimates the maximum cross section reach by up to 15\% (depending on DM mass) in comparison to the improved SGED\textbf{/}KS treatment of the Appendix, i.e. the improved calculation leads to a more limited exclusion range than claimed by KS, making their approximation not conservative.
\item SGED and KS take $L = z_{det}$, the detector depth.  This is clearly inaccurate for the ensemble of DM particles,  since it is true for only an infinitesimal fraction of trajectories which are incident vertically and suffer no scatterings, or scatter purely in the forward direction.  %If average trajectories could be described as an isotropic random-walk in 3 dimensions, then the rms displacement after $N$ scatterings is $\sqrt{N} \lambda$ where $\lambda$ is the step length, but the pathlength traveled is   The mean displacement in any specified direction is $\\sqrt{frac{N}{3}} \lambda$.  Thus to reach depth $z_{det}$ requires on average $N = 3 (z_{det}/\lambda)^2$ scatters, for a mean total distance traveled  and total pathlength $3 \frac{z_{det}^2}{\lambda}$ \sim \frac{z_{det}}{\lambda}$ and the distance traveled would be $L \sim  \frac{z_{det}^2}{\lambda}  \gg z_{det}$.  
Figures \ref{pathlengthE_D} and \ref{pathlengthPb_D} show the actual distribution of path lengths of successful particles as determined by Monte-Carlo simulations, for 3 choices of DM mass. The vertical red lines correspond to the thickness of each of these shielding layers above the DAMIC detector. Successful DM particles travel on average a larger total path length in the Earth's crust by a factor of 1.1 -- 1.23, and in the lead shield by a factor of 1.6 -- 1.9, depending on the DM mass. If these larger values of $L$ are substituted in eq.~\eqref{bound}, it reduces $\sigma_{\rm max}$, i.e. the ``improved'' calculation leads to a more limited exclusion range than claimed by KS, again making their approximation not conservative.
In fact, simply using a Monte-Carlo average-path length instead of the linear vertical distance traveled, is not valid either, and does not improve the accuracy of eq.~\eqref{bound}, as we see in the following. 
%\iffalse
\begin{figure}[tbp]\centering
	\begin{subfigure}{.495\textwidth}
		\centering
		\includegraphics[width=\linewidth]{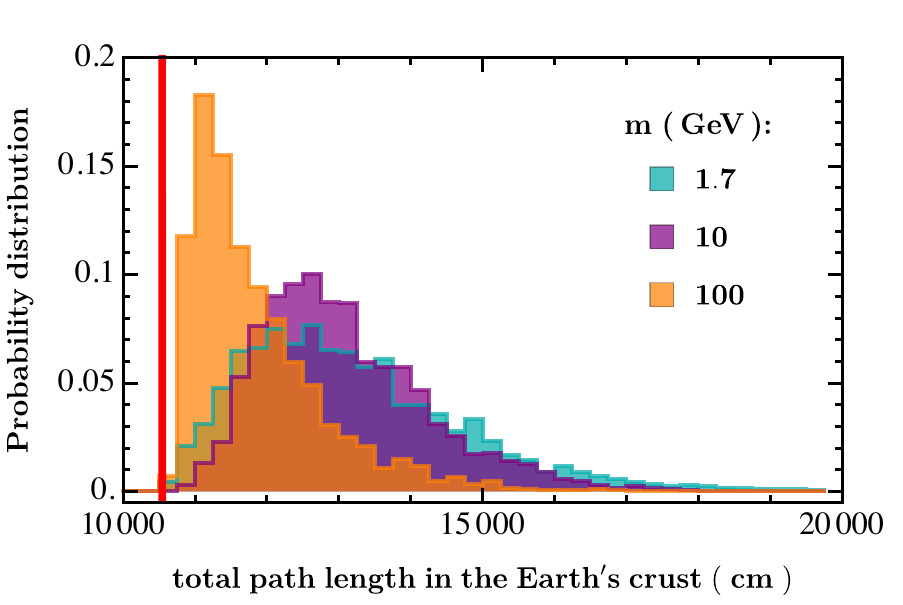}
		\caption{}
		\label{pathlengthE_D}
	\end{subfigure}\hspace{0.01cm}
	\begin{subfigure}{.495\textwidth}
		\centering
		\includegraphics[width=\linewidth]{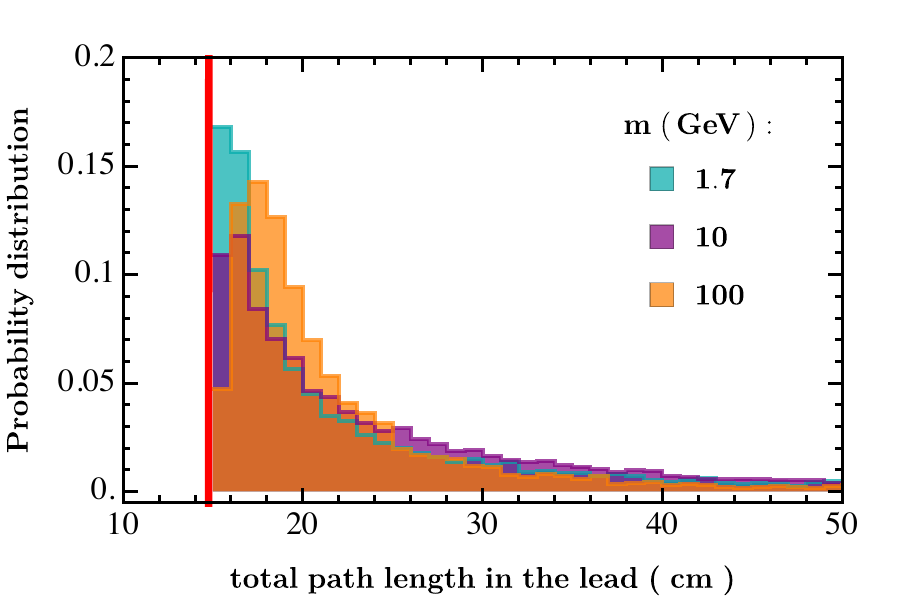}
		\caption{}
		\label{pathlengthPb_D}
	\end{subfigure}
	\caption{The distributions of the total path length for successful DM particles for masses 1.7, 10, and 100 GeV when passing through (a) the Earth's crust and (b) the lead shield.  The SGED approximation is shown by the vertical red lines at 10670 and 15 cm.}
	\label{pathlengthD}
\end{figure}

\item
The SGED\textbf{/}KS treatment does not provide a means to incorporate the momentum transfer dependence of the DM-nucleus cross section, i.e., the form factor.  As the mass of the DM particles increases, the momentum transfer increases and the form factor deviates more from unity. The impact of a form factor is always to increase $\sigma_{\rm max}$, since the form factor is $\leq 1$.  The magnitude of the effect can only be judged with Monte-Carlo simulations.  We find that the correct value of $\sigma_{\rm max}$ for the DAMIC experiment accounting for the form factor, is 16\% higher for $m=100$ GeV DM, than if the form factor is set to 1.

\item 
A more serious, conceptual error arises in the SGED/KS value of $\langle f\rangle$. In most models, DM-nucleon scattering is isotropic in the CM frame.  Therefore in calculating $\langle f\rangle$ from eq.~\eqref{DeltE}, SGED/KS take $\langle 1-\cos\,\xi_{CM}\rangle = 1$.  However the distribution of scattering angles of \emph{successful} DM particles off nuclei in the Earth's crust and the lead shield is not representative of average particles;  it is far from isotropic and favors forward scattering. This is for two reasons:  small angle scattering reduces the energy loss, as well as shortening the total path length.  Figures \ref{xiE_D} and \ref{xiPb_D} show the distributions of the scattering angles in the CM frame, of all the scatterings of successful DM particles in the Earth's crust and the lead shield. Although the scatterings of the ensemble of DM particles are isotropic in the CM frame, the scatterings of successful DM particles in the CM frame are not isotropic. Instead, they are skewed toward forward scattering which lowers the energy-loss. This works in the direction of making the true cross section reach stronger in comparison to the SGED\textbf{/}KS value, because \emph{successful} DM particles lose on average a smaller fraction of their energy in each scattering in the Earth's crust and in the lead shield than does an average DM particle\footnote{We rather arbitrarily have chosen to characterize successful rather than capable particles in the plots of this section.  Both classes behave similarly, but successful particles on average have somewhat higher energies when reaching the detector than capable ones.}. 

\begin{figure}[tbp]\centering
	\begin{subfigure}{.495\textwidth}
		\centering
		\includegraphics[width=\linewidth]{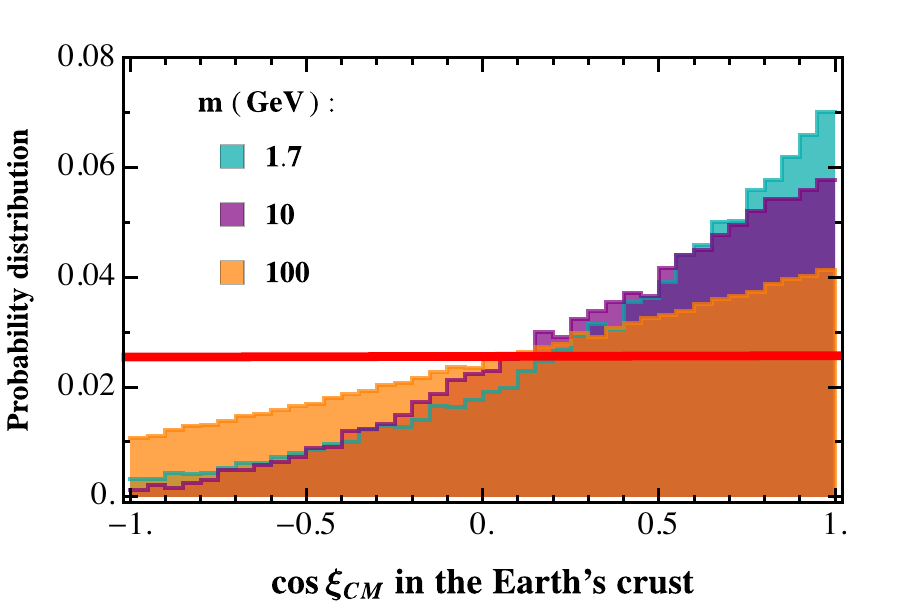}
		\caption{}
		\label{xiE_D}
	\end{subfigure}\hspace{.01cm}
	\begin{subfigure}{.495\textwidth}
		\centering
		\includegraphics[width=\linewidth]{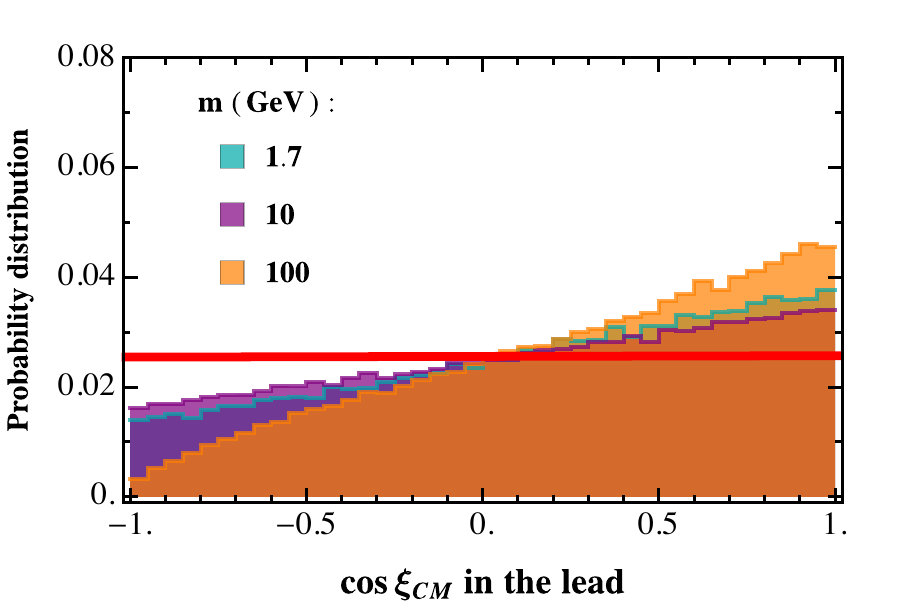}
		\caption{}
		\label{xiPb_D}
	\end{subfigure}
	\caption{The distributions of the CM scattering angles of all scatterings in the trajectories of successful DM particles for masses 1.7, 10, and 100 GeV when passing through (a) the Earth's crust and (b) the lead shield. The solid line corresponds to the distribution of the CM scattering angles of all particles, successful or not.}
	\label{xiD}
\end{figure}

\item The other most important quantity in the energy loss estimate is $N$, the average number of scatterings, which is $L/\lambda$ in the SGED/KS formulation where $\lambda$ is taken interchangeably to be the average path length between scatterings and the interaction length, $\lambda = (n\,\sigma)^{-1}$.  As noted, taking $L \rightarrow z_{det}$ is not valid.  Less obviously incorrect, but in fact an important source of error, is identifying the average path-length between scatterings as being the interaction length $(n\,\sigma)^{-1}$.  Figure~\ref{nmD} shows the distributions of the number of scatterings in the Earth's crust and the lead shield for successful DM particles. The arrows indicate the number of scatterings averaged over all trajectories (including unsuccessful ones) for particles propagating a distance $z_{det}$, i.e., $\frac{z_{det}}{\lambda_{eff}}$ in the Earth's crust and $\frac{z_{Pb}}{\lambda_{Pb}}$ in the lead shield\footnote{In a mixed-composition medium like the Earth's crust, $\lambda_{eff}$ is the effective scattering length;  see the Appendix.}.  These are the values that are used in the SGED/KS treatment.  The values of $<N>$ using the true path length instead of $z_{det}$ are of course even higher: for $m = \{$1.7, 10, 100$\}$ GeV respectively, the average number of scatterings are $\{$19.2, 23.4, 20.5$\}$ in the Earth and $\{$1.67, 3.5, 17.4$\}$ in the lead. These are dramatically larger than the peak values of the true distribution for successful particles, which range from 4-7 in the Earth and 0-2 in the lead.
Figure~\ref{nE_mD} shows that the average path length between scatterings in the Earth's crust of successful DM particles is 2.7 to 3.4 times larger than the mean scattering length, $ \lambda_{eff} $, depending on the DM mass. 

\begin{figure}[tbp]\centering
	\begin{subfigure}{.495\textwidth}
		\centering
		\includegraphics[width=\linewidth]{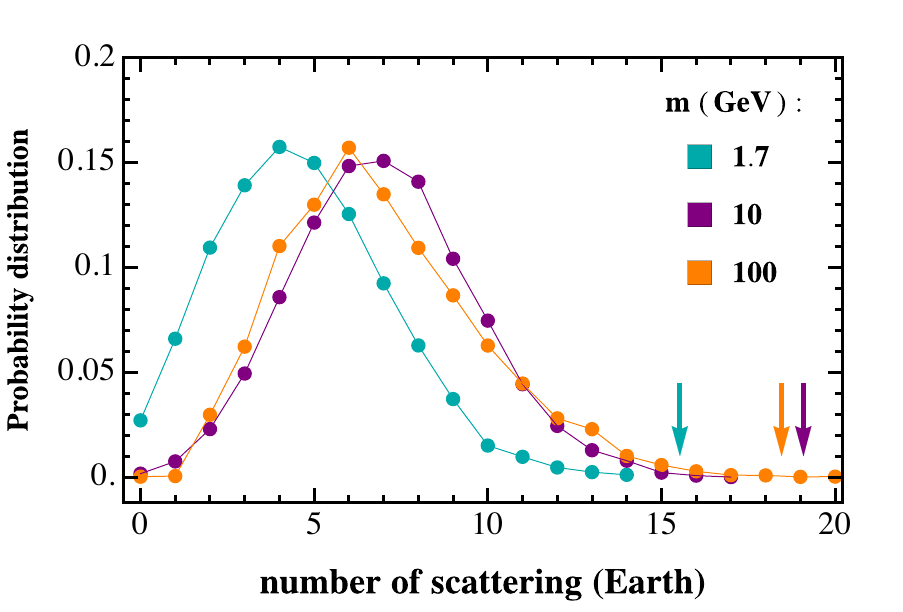}
		\caption{}
		\label{nE_mD}
	\end{subfigure}\hspace{.01cm}
	\begin{subfigure}{.495\textwidth}
		\centering
		\includegraphics[width=\linewidth]{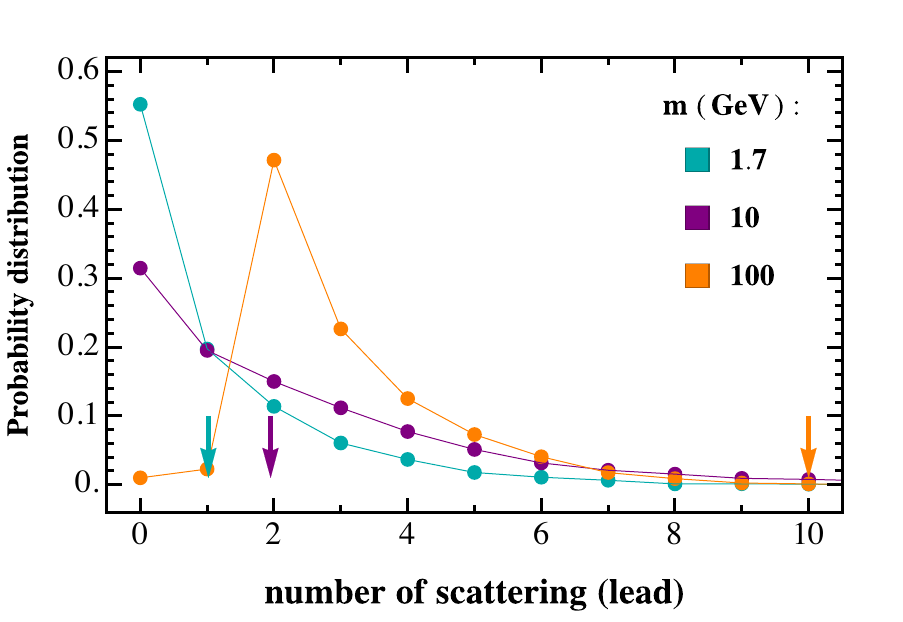}
		\caption{}
		\label{nPb_mD}
	\end{subfigure}
	\caption{The distributions of the number of scatterings for successful DM particles for masses 1.7, 10, and 100 GeV when passing through (a) the Earth's crust and (b) the lead shield. Arrows show $\frac{z_{det}}{\lambda_{eff}}$ and $\frac{z_{Pb}}{\lambda_{Pb}}$ for these DM masses. These are the values which are used in the SGED\textbf{/}KS approximation; the large departure from the actual means of the distributions explains the failure of this approximation.}
	\label{nmD}
\end{figure}
\end{enumerate}

Taken together, successful DM particles are much more abundant at the detector than would be expected from the SGED\textbf{/}KS approximation, due to exploiting the tails of scattering length and scattering angle distributions. The fallacy in using the mean approximation derives from the fact that minimizing the energy loss enhances the probability of triggering the detector. %This makes Monte-Carlo cross section bounds stronger by a factor of 2.7 -- 3.4 in comparison to the SGED\textbf{/}KS bounds, as successful DM particles travel 2.7 -- 3.4 times further than average between consecutive scatterings.
\par
\begin{figure}[tbp]\centering
	\includegraphics[width=\textwidth]{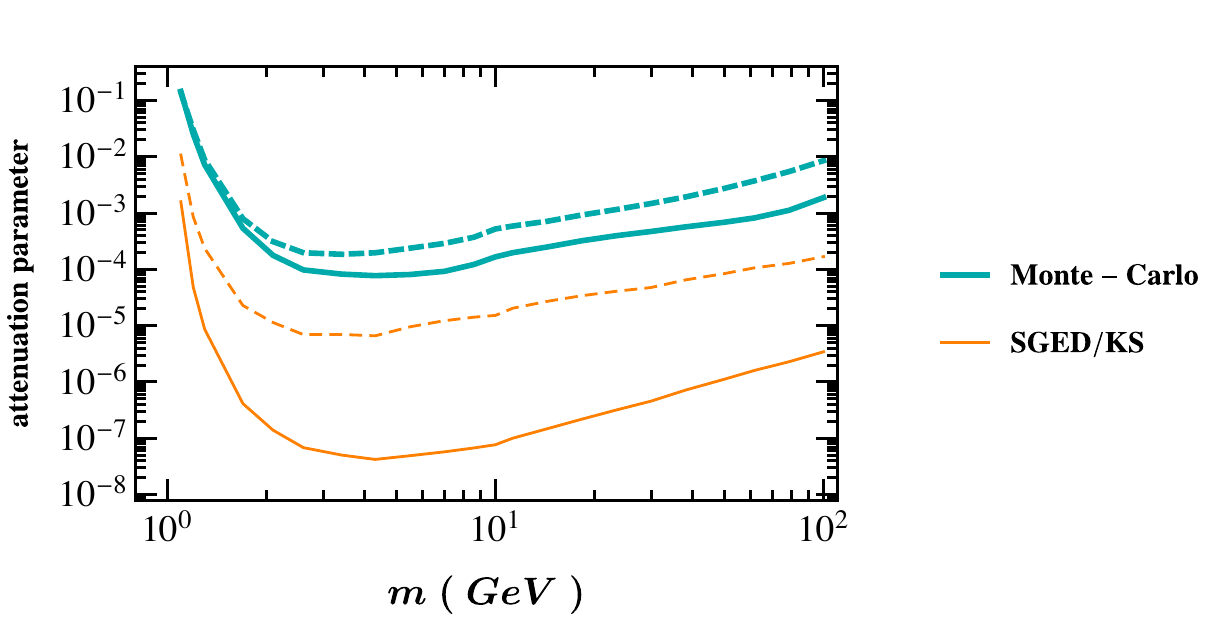}
	%\caption{\label{DAMIC_att@sigma_SGED}The ratio of the number of events at DAMIC's depth to the number of events if the detector was on the Earth's surface (solid lines) and the ratio of DM particles that can potentially trigger the detector to the total number of DM particles on the Earth's surface (dashed lines), for the limiting cross section of the SGED\textbf{/}KS method, using the Monte-Carlo simulation (cyan) and the the SGED\textbf{/}KS method.}
	\caption{\label{DAMIC_att@sigma_SGED}The attenuation parameters for the total number of detected events (solid) and the number of \emph{capable} DM particles (dashed), calculated with the Monte-Carlo (cyan, solid) and the improved SGED\textbf{/}KS (orange, thin solid) methods, for the limiting cross section of the improved SGED\textbf{/}KS method.}
\end{figure}

The net effect of the inaccuracies of the SGED\textbf{/}KS description is to underestimate the number of detected events by up to a factor $10^3$, for the $\sigma_{\rm max}$ obtained by KS.  Figure~\ref{DAMIC_att@sigma_SGED} shows the attenuation parameters for the total number of events (solid lines) and number of \emph{capable} DM particles (dashed lines) using the Monte-Carlo (thick) and the improved SGED\textbf{/}KS methods (thin, calculated as described in the Appendix) for the limiting cross section of the improved SGED\textbf{/}KS method. This difference corresponds to as much as a three order-of-magnitude larger number of events at the limiting cross section, for DM mass $\sim$10 GeV. This effect is greatest at 10 GeV due to the similarity of the DM and the target nuclei masses and hence maximization of the DM energy loss per collision.

\par
\begin{figure}[tbp]\centering
	\includegraphics[width=.9\textwidth]{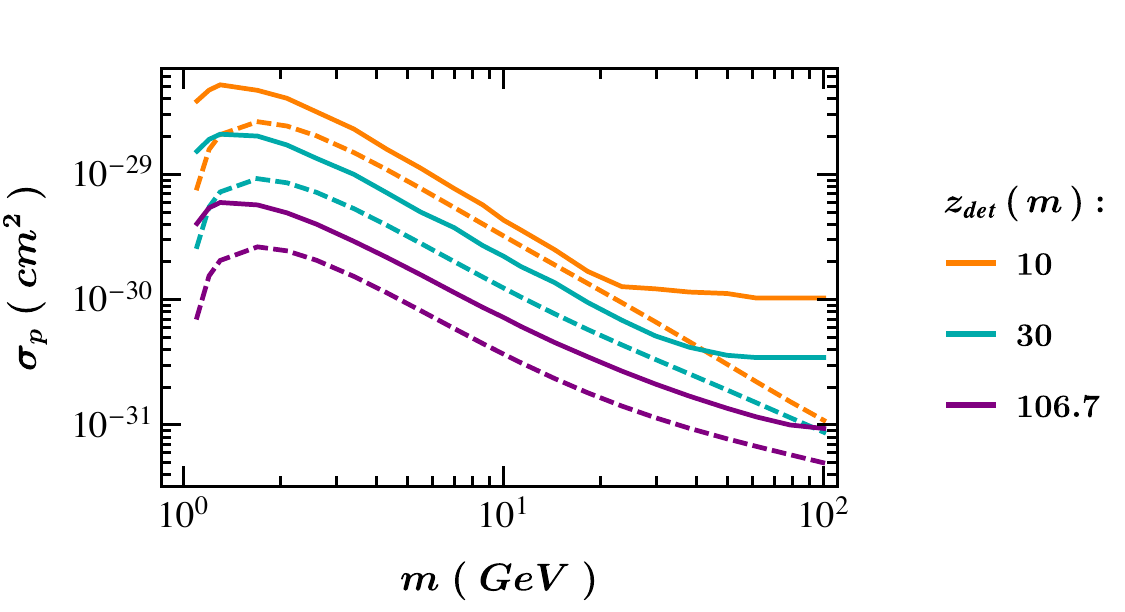}
	\caption{\label{DAMIC_3depth}The 90\% CL DM-proton cross section lower bounds deduced by the importance sampling Monte-Carlo simulation (solid lines) and the SGED\textbf{/}KS method (dashed lines) for the three detector depths of 10, 30, and 106.7 meters underground.}
\end{figure}

Figure~\ref{DAMIC_3depth} shows the 90$\%$ CL DM-proton cross section reach, $\sigma_{\rm max}$, using the Monte-Carlo simulation (solid lines) and the SGED\textbf{/}KS method (dashed lines), taking -- hypothetically -- DAMIC's results to have been obtained at three different detector depths (10, 30, and 106.7 meters underground), to explore how the discrepancy depends on depth.  At the actual 106.7m depth, the SGED\textbf{/}KS method underestimates the true cross section reach of DAMIC by a factor of 1.8 -- 5.6. For smaller DAMIC detector depths, the discrepancy between the true result and the result of the SGED\textbf{/}KS method becomes even larger for large DM masses. 

\section{Regime of applicability of the SGED/KS approach}
\label{EKS}

Estimating the maximum cross section reach of DAMIC to a factor-few accuracy might seem sufficient, but the KS estimate of \sigmax\ for DAMIC left a window for moderately-interacting DM ($\sigma_p \sim \mu$b) in the $\sim$GeV mass range. We closed this window through our accurate Monte Carlo analysis based on importance sampling~\cite{bounds}. One can ask if the SGED/KS type of analysis can be accurate in other contexts, for instance when the energy loss per collision is small, so that accumulated favorable fluctuations in the the trajectories may have less impact.

Emken, Kouvaris and Shoemaker~\cite{Emken} (EKS) have recently investigated the SGED/KS approach to take account of energy loss via scattering on nuclei in the Earth's crust, in the context of constraining the dark photon model with DAMIC and XENON10. In the small mass range, $m\leq $ GeV, the energy loss per interaction is small, so many interactions are needed to lose an important amount of energy, and there is no suppression due to form factors.  In this regime the number of interactions must be large before energy loss in the overburden has any effect, so the continuous energy loss approximation \emph{per se} should be good. Using the correct mean path-length and not the linear-vertical-distance assumption is of course important, and EKS~\cite{Emken} performed a Monte Carlo simulation of trajectories.  

EKS compared Monte-Carlo and analytic results for the total energy loss after a fixed total path-length $d$, as a function of mass, in their Fig. 3~\cite{Emken}. They report that the analytic approximation is ``remarkably good over a wide range of masses''.   While true --- indeed, guaranteed by the central-limit theorem --- this agreement does not bear on the validity of using the SGED/KS approximation for determining the cross section reach of some experiment as discussed in eqs~\eqref{EfEi} and~\eqref{SGEDESimple} and in more detail in the Appendix. A simulation is required to know the mean value of the energy loss per collision and mean number of collisions of \emph{capable} particles \emph{at the detector} -- not the mean values for all particles after a fixed total path length $d$. 

Although EKS simulate trajectories, they use a seemingly ad-hoc shortcut to find \sigmax,  c.f., their equation (16)~\cite{Emken}. They introduce a ``critical cross section'', defined to be the cross section such that the mean DM speed at the detector depth, $\langle v\rangle$, is at least $5\,\Delta v$ below the minimum speed to produce a signal in the detector, where $\Delta v$ is the standard deviation in the arrival speeds at the detector, given the assumed starting DM speed. This $5\,\Delta v$ prescription is not demonstrated to be correct. Its rationale is probably that if the distribution of final velocities is Gaussian, then the tail above $5\,\Delta v$ contains a fraction $\rm{2.8\times10^{-7}}$ of the distribution. A high statistics simulation would be required to demonstrate that the distribution is Gaussian or that specifically the $5\,\Delta v$ prescription assures the correct attenuation.%(See point 1 in our discussion of the previous section).  

\begin{figure}[tbp]\centering
	\includegraphics[width=.99\textwidth]{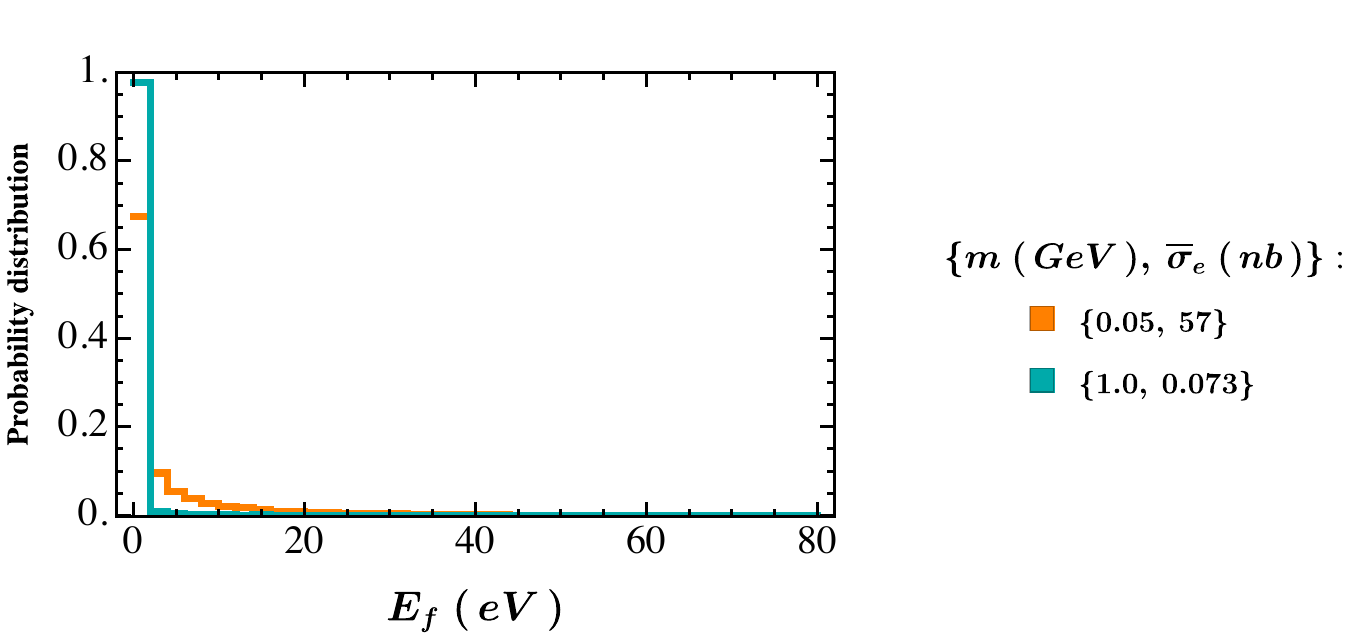}
	\llap{\shortstack{%
			\includegraphics[scale=.33]{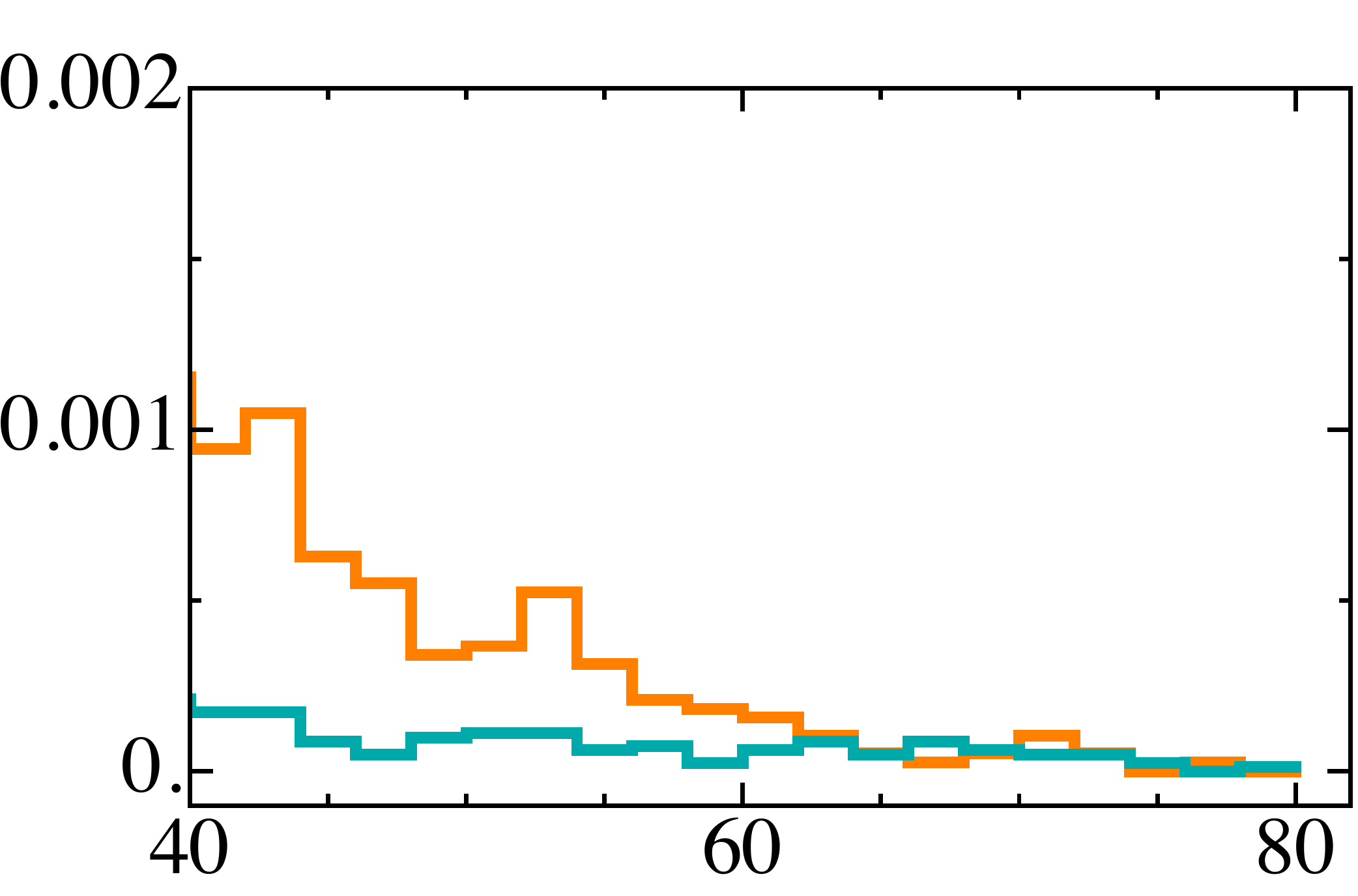}\\
			\rule{0ex}{.7in}%
		}
		\rule{2.45in}{0ex}}
	\caption{\label{EKS_spectrum}The final energy distribution of capable DM particles at DAMIC's depth for the EKS critical DM-electron cross section. The inset figure is the zoom-in view of the spectrum for DM particles with $E_f\geq40$ eV. The fraction of DM particles reaching DAMIC which are above the DAMIC threshold of 40 eV using the EKS critical cross section is: $\rm{5.7\times10^{-3}}$ for 50 MeV (orange) and $\rm{2.2\times10^{-3}}$ for 1 GeV (cyan). }
\end{figure}

We tested the EKS procedure by using importance sampling to perform a full simulation of the expected spectrum in DAMIC, for the critical cross sections found by EKS for two DM masses: 50 MeV and 1 GeV. Figure~\ref{EKS_spectrum} shows the final energy spectrum of capable DM particles at DAMIC's depth that we found using the EKS critical cross sections. Had the EKS result been correct, we should have obtained 106 events for both masses, but instead we found $\rm{4.5\times10^{8}}$ events for 50 MeV and $\rm{2.5\times10^{6}}$ events for 1 GeV. 
\par 
To determine the correct prescription to replace the EKS ``$5\,\Delta v$'' criterion requires a much higher-statistics simulation, with $> 10^8$ particles for each mass and cross section combination, to measure the high-speed tail of the distribution. In the end, that amounts to performing the full, high statistics simulation. Perhaps in the small energy loss case, regularities can be found to avoid simulating for the full, fine grid of cross section and masses.

\section{Importance Sampling}~\label{IS}
\begin{figure}[tbp]\centering
	\includegraphics[width=0.9\linewidth]{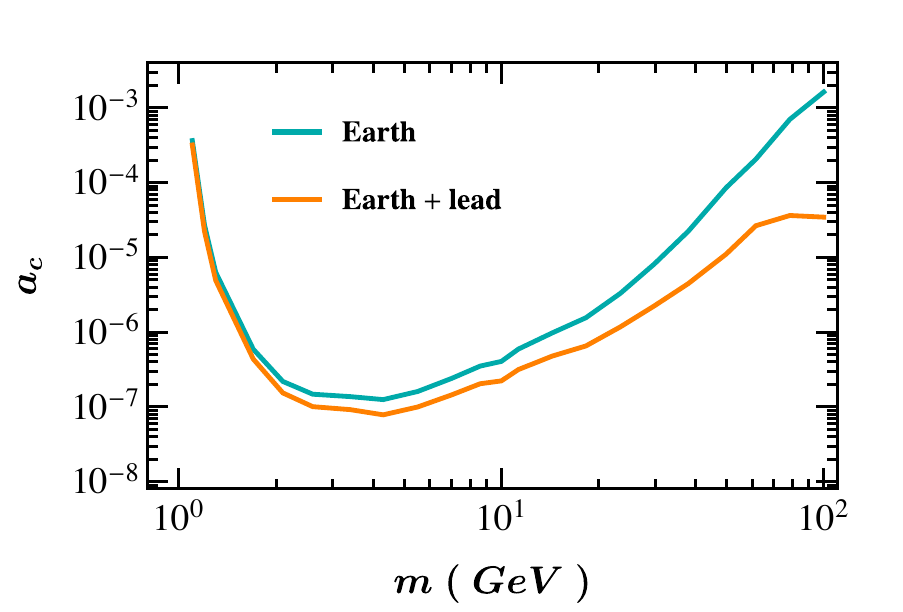}
	\caption{\label{DAMIC_att_c@sigma_MC}  The attenuation parameter $a_c$ of capable DM particles after passing through 106.7 meters of the Earth's crust and 6 inches of the lead shield for the limiting cross section from DAMIC found by the Monte-Carlo method.}
\end{figure}

In this section, we develop an importance sampling technique which makes accurate Monte-Carlo simulation of DM attenuation computationally feasible. As noted in~\cite{bounds} and shown in figure~\ref{DAMIC_att_c@sigma_MC}, the attenuation parameter for capable DM particles, $a_c$, can be as small as $\rm{8\times10^{-8}}$ for DM mass 4 GeV, at the DAMIC limiting cross section and depth. This means that the propagation of $\rm{1.25\times10^{7}}$ particles with this mass must be simulated on average, to achieve one DM particle with enough energy at the DAMIC surface to potentially trigger the detector, and of course many more trajectories must be simulated to get a statistically adequate prediction for the spectrum of capable particles at the detector. As experimental sensitivities increase, ever-higher attenuation factors will be encountered in the analyses, until a positive DM signal is seen, at which point higher statistics simulations will be motivated, for more detailed interpretation of the signal.
\par
The immediate goal motivating this study, was to determine $\tilde{f}(v)$, the normalized speed distribution of capable DM particles, and $a_c(\sigma_p)$, the attenuation parameter.  $a_c(\sigma_p)$ is the ratio of the number of capable DM particles at the detector at DAMIC's detector depth, $z_{det}=106.7$ meters, to the total number of capable DM particles hitting the Earth's surface over the course of the experiment, for the assumed $\sigma_p$.  %[I think we discussed this, but just to be triply sure: are you sure of this definition of $a_c$, and that it isn't defined to be the ratio of capable particles after the overburden relative to the number at the surface of the overburden?  The English meaning of the word attenuation would imply that the latter should be the definition.] 
Knowing $\tilde{f}(v)$ and $a_c(\sigma_p)$ enables us to calculate the expected total number of events for DAMIC's silicon target of mass number $A=$ 28, \cite{bounds}:
\beq\begin{split}
	\label{eDAMIC_N}
	N=&\,t_e\,M_A\,A^2\,\eta(m)\,\frac{a_c(\sigma_p)\,\rho\,\sigma_p}{4\,m\,\mu^2_p}\int_{0.55\,keV}^{7\,keV}F_A^2(E_r)\,dE_r\int_{v_{min}(E_r,A)} v\,\tilde{f}(v)\,dv\;,
\end{split}\eeq
where $\rho$ is the DM local mass density, $\sigma_p$ is the DM-proton cross section and $F_A(E_r)$ is the nuclear form factor given in eq.~\eqref{e_FormFactor}. $\eta(m)$ is the fraction of capable DM particles at the Earth's surface.\footnote{In principle, we could combine the $a_c(\sigma_p)$ and $\eta(m)$ into a single variable. But, we choose to keep these two separate to be able to compare the population of capable DM particles above and below the detector by $a_c(\sigma_p)$.} The DAMIC detector accumulated a total exposure of $t_e\,M_A=$ 107 g$\cdot$days from June 2010 to May 2011, and observed a total of 106 events with an ionization signal between 40 $\rm{eV_{ee}}$ (corresponds to nuclear recoil energy 0.55 keV) and 2 k$\rm{eV_{ee}}$ (corresponds to nuclear recoil energy 7 keV)~\cite{Barreto2012264, bounds}. $v_{min}$ is the minimum speed that a DM particle on the detector needs in order to deposit the recoil energy $E_{r}$ into the detector (see eq.~\eqref{e_v'm}).
\par
The details of the Monte-Carlo simulation of trajectories in the overburden are spelled out in \cite{bounds}. To increase the efficiency of the simulation we wish to increase the portion of simulation time devoted to trajectories which can potentially produce an event in the detector, but maintain the correct relative weight between different types of trajectories in order to be able to calculate any physical quantity of interest. Such a strategy is an instance of importance sampling~\cite{Bucklew_IS,Owen_IS}.
\par
A key determinant of the energy loss during propagation is the number of interactions, as evident from eq.~\eqref{EfEi}. Fewer interactions means less energy loss, so increasing the likelihood of a longer-than-average distance to the next interaction (step length), increases the fraction of successful events.  This biasing of the simulation must be accounted for with a weight factor, as detailed below. 
\par
Another way to reduce the total energy loss is to reduce the mean fractional energy loss per collision.  This can be done by enhancing the likelihood of forward scattering, i.e., modifying the CM scattering angle distribution to have a mild slope, to resemble the distribution in figure \ref{xiD}. The procedure for implementing this modification is directly parallel to that for modifying the step length distribution.  In the following we focus on the step length modification, to explain the method.  That provided sufficient increase in efficiency for our DAMIC analysis reported in~\cite{bounds};  the procedure for modifying the angular distribution is perfectly analogous.
\par
The probability of step length between $r$ and $r + dr$, is $p(r) \, dr$ where
\beq
\label{eW(r)}
p(r)=\frac{1}{\lambda_{eff}}\,e^{-r\,/\lambda_{eff}}\;,
\eeq
and $\lambda_{eff}\equiv(\sum_{A}n_A\sigma_A)^{-1}$ is the mean interaction length in the overburden with a mix of nuclei. To increase the sample size of $capable$ DM particles reaching the detector, we modify the step length distribution of DM particles to 
\beq
\label{eW'(r)}
q(r)=\frac{1}{(1+\delta) \lambda_{eff}}\,e^{-r\,/(1+\delta) \lambda_{eff}}\;.
\eeq
Taking $\delta>0$ shifts the average step length from $\lambda_{eff}$ to $(1+\delta)\lambda_{eff}$ while retaining the qualitative character of the step length distribution.    
\par
To correct for not sampling from the true path length distribution for the $j$-th DM particle, we assign a weight factor for each
step $i$ along its trajectory: $w_{ij}=\frac{p(r_{i})}{q(r_{i})}$.  There is one more weight factor than interactions because a step length must be chosen for the initial particle before the first interaction and for every subsequent step.  %It is only retrospectively that the last interaction is recognized as being last and the process is terminated, when the depth is found to be below $z_{det}$.   
We thus label the incident particle before scattering as $0$, so the weight factor associated with propagation to the first interaction point is $w_{0j}$.   The total weight factor for the $j$-th particle is then
\beq
\label{ew}
w_j=\prod_{i=0}^{n_j}w_{ij}=\prod_{i=0}^{n_j}\frac{p(r_{i})}{q(r_{i})}\;,
\eeq
where $n_j$ is the number of scatterings of the $j$-th particle in the overburden.  %For the last step during which $z_{det}$ is passed, $r_i$ is the chosen step length, not the 
\par
It is essential to choose the modified distribution $q(r)$, such that there is no part of the parameter space $r$ for which there is an extremely large weight factor.  If this is not respected, higher statistics sampling is required to smooth out fluctuations from occasional trajectories having very large weight factors.  In particular, $q(r)$ should not have a cut-off; i.e., it should not vanish in any region where $p(r)$ is finite.
\par
The importance sampling estimator of $f$ --- where $f$ is any physical quantity of interest, for instance the normalized speed distribution of capable particles --- is calculated as it would be without the weighting procedure, but with the contribution of the $j$-th particle weighted by $w_j$.   Symbolically,  
\beq\begin{split}
	\label{egq2}
	\hat{f}=\frac{\mathlarger{\sum}\nolimits_{j=1}^{N} f_j\,w_j}{\mathlarger{\sum}\nolimits_{j=1}^{N} w_j}\;,
%	\hat{f}_{q,N}=\frac{\mathlarger{\sum}\nolimits_{j=1}^N f_j\,w_j}{\mathlarger{\sum}\nolimits_{j=1}^N w_j}\;,
\end{split}\eeq
where $N$ is the total number of relevant particles in the simulation.   
\par 
In order to calculate physical quantities, one has to know the weights of every particle in the ensemble of contributing particles.                                         
If we are only interested in capable particles, we need not follow particles which reflect or fall below the minimum energy before reaching the detector, apart from recording them in the total number simulated.  Some examples will clarify.
\par
Consider first the case of unit-normalized distributions of capable particles such as $\tilde{f}(v)$. Then, $N$ in eq. \eqref{egq2} is the number of capable particles. For instance, the unit-normalized speed distribution of capable particles is found by summing the total weight of capable particles in each speed bin and dividing by the total weight of capable particles simulated.   %\par
We can also find the value of some physical quantity concerning each step of DM trajectories, e.g., the CM scattering angle distribution of all the scatterings (see figure~\ref{xiD}) in trajectories of successful DM particles.  Calling the observable $g$,
\beq\begin{split}
	\label{egq3}
	\hat{g}=\frac{\mathlarger{\sum}\nolimits_{j=1}^{N_s}\mathlarger{\sum}\nolimits_{i=0}^{n_j} g_{ij}\,w_{ij}}{\mathlarger{\sum}\nolimits_{j=1}^{N_s}\mathlarger{\sum}\nolimits_{i=0}^{n_j} w_{ij}}\;,
\end{split}\eeq
where $w_{ij}$ is the weight of $j$-th DM particle in $i$-th 
step of its trajectory and $N_s$ is the total number of particles in the ensemble, in the example, successful DM particles.
\par
Computation of absolutely normalized quantities such as the reflected fraction or the attenuated fraction of DM particles, is equally straightforward because the mean weight of all particles is 1 by construction, as follows from eq.~\eqref{ew}, and readily verified for the path length re-weighting of eq.~\eqref{eW'(r)}. Thus the total weight is simply the number of particles simulated, $N_{tot}$. Let us use $a_c$ as an example.  $a_c$ is defined as the ratio of capable particles after the overburden relative to the number at the surface of the overburden. Designating the total number of incident particles above the threshold and the total number of capable DM particles by $N_{tot}$ and $N_c$, respectively, the attenuation parameter of capable particles is
\beq\begin{split}
	\label{ac}
	a_c =\frac{\mathlarger{\sum}\nolimits_{j=1}^{N_c} w_j}{N_{tot}}\;.
\end{split}\eeq
This is because the sum of weights of all particles is $N_{tot}$, so the particles \emph{not contributing} to the quantity of interest do not have to be followed because their weights are not required.  

Features of the non-capable particles can also be calculated, for instance some property of incident particles which are reflected.  In this example, the weights of whatever particles are of interest (e.g., the ones which reflect) must be calculated, because their weights are needed in~\eqref{egq2}, but the weights of populations which are not of interest are not needed since the total weights sum to $N_{tot}$ and individual weights of particles not contributing to the observable are not needed.
\par
A word is in order about the last step, when the particle passes $z_{det}$ and reaches the detector, or more generally any discontinuity in $\lambda_{eff}$.   One could imagine that some modification of the weighting procedure might be needed, but it can be seen as follows that no modification is needed.  For each generated step, one checks to see if the particle is still in the first medium at the end of the step, e.g., if $z \leq z_{det}$ or not. If it is not still in the first medium, it is put into the ensemble of particles (capable ones, in our case, if its energy is sufficient) which passes into the next material (the detector, for instance). Then, the interaction of that population with the next material is calculated. The fact that the generated step length would place it some distance into the next material is irrelevant because its likelihood to interact in a given distance interval going forward is independent of its past behavior. The crucial point is that if $L$ is the remaining distance along the trajectory from a given position to the detector depth, then a fraction $e^{-L/\lambda}$ will reach the detector without another interaction, and this fraction is preserved in the weighted distribution, as required and guaranteed by the definition of the weights. 
\par
For $\delta=$ 0.4, 0.6, 0.8, the sample size of capable DM particles for a given number of simulated events increases by a factor of 100, 400, 1000 respectively relative to the brute-force simulation with $\delta=$0. This enables our Monte-Carlo simulation to simulate the DM velocity distribution on the surface of the DAMIC detector to a specified statistical accuracy using 100 -- 1000 times less computational resources, by wasting less computational power in simulating DM particles which end up losing too much energy in the Earth's crust or the lead shield. The DM{\scriptsize ATIS} code, which we developed for this study, will be made publicly available on-line~\cite{code}.  

\section{Validation of importance sampling}~\label{valid}
In this section, we compare a number of physical distributions simulated with the brute-force and the importance sampling Monte-Carlo methods.  They are found to be indistinguishable, thereby validating the importance sampling method.  At the same time, we learn what are the actual characteristics of the trajectories of capable DM particles.  This is useful for developing intuition and building an even more computationally efficient re-weighting method. We focus on DM mass 1.7 GeV.  This is not only a specifically interesting mass~\cite{Farrar2003,Farrar:2003qy,Farrar:2004qy}, but also has nearly maximal attenuation (see figure~\ref{DAMIC_att_c@sigma_MC}).  We set the DM-proton cross section to DAMIC's sensitivity-limit value for this DM mass, $\sigma_p=$ 5.7$\times\rm{10^{-30}\,cm^2}$ (see figure~\ref{DAMIC_3depth}) \cite{bounds}. 
\par
\begin{figure}[tbp]\centering
	\begin{subfigure}{.495\textwidth}
		\centering
		\includegraphics[width=\linewidth]{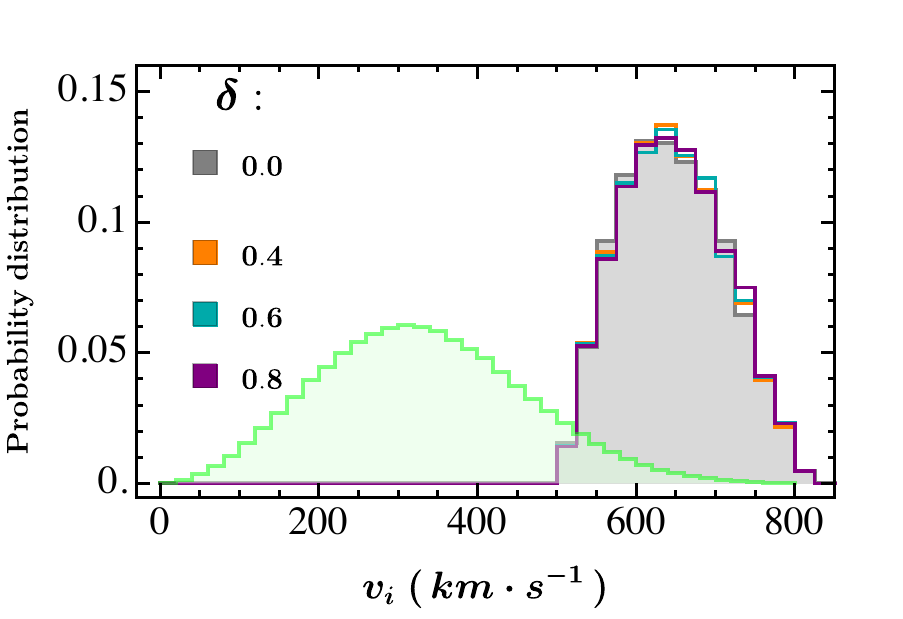}
		\caption{}
		\label{vi_D}
	\end{subfigure}\hspace{0.01cm}
	\begin{subfigure}{.495\textwidth}
		\centering
		\includegraphics[width=\linewidth]{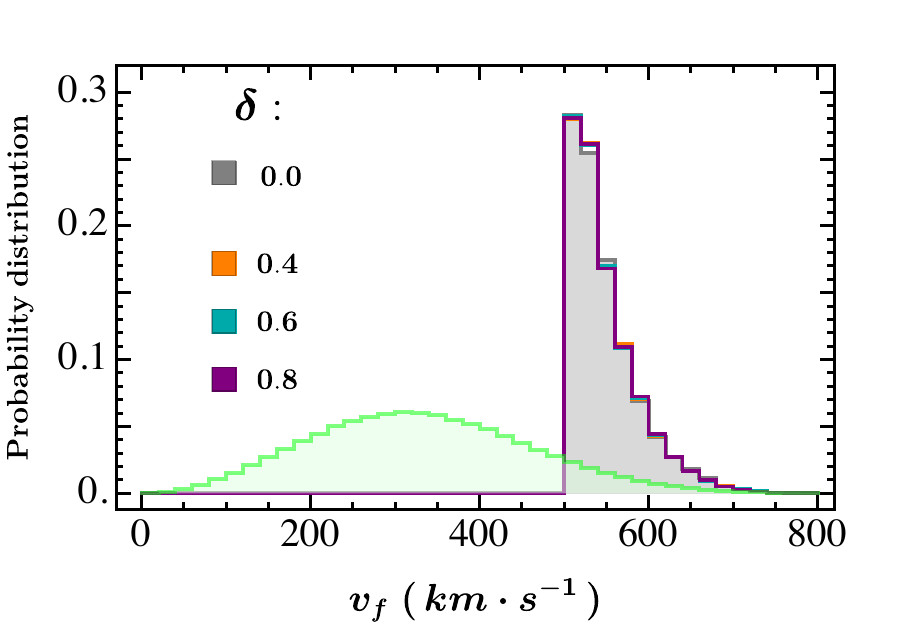}
		\caption{}
		\label{vf_D}
	\end{subfigure}
	\caption{(a) The initial (i.e., at Earth surface) and (b) the final DM speed distributions of capable DM particles for our four choices of $\delta$, for DM mass of 1.7 GeV and a DM-proton cross section of 5.7$\times\rm{10^{-30}\,cm^2}$. The light-green histogram shows the speed distribution of all of DM particles on the Earth's surface. All the distributions are normalized to 1. There is no significant difference in the distribution between the brute-force ($\delta=0$) and the importance sampling results.}
	\label{f_speedD}
\end{figure}
\begin{figure}[tbp]\centering
	\begin{subfigure}{.495\textwidth}
		\centering
		\includegraphics[width=\linewidth]{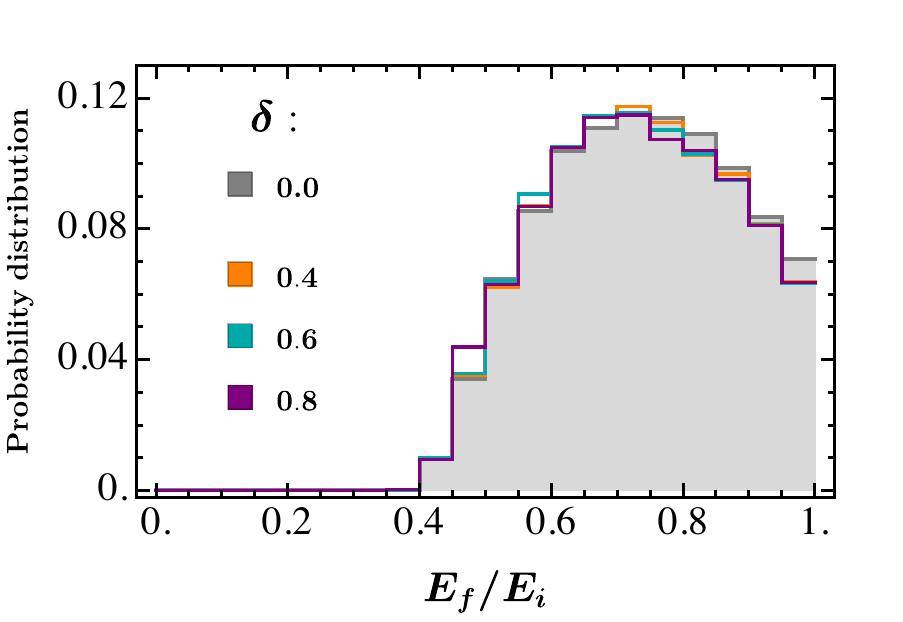}
		\caption{}
		\label{EfEi_D}
	\end{subfigure}\hspace{0.01cm}
	\begin{subfigure}{.495\textwidth}
		\centering
		\includegraphics[width=\linewidth]{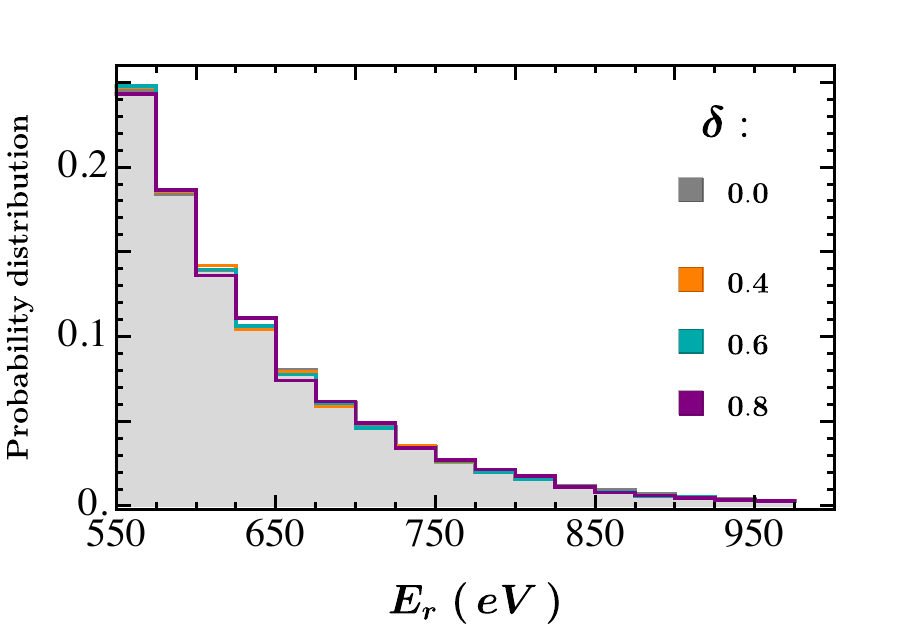}
		\caption{}
		\label{Er_D}
	\end{subfigure}
	\caption{The distributions of (a) the ratio of final to initial energy and (b) the nuclear recoil energy in the detector of capable DM particles, for DM mass of 1.7 GeV and a DM-proton cross section of 5.7$\times\rm{10^{-30}\,cm^2}$. 550 eV is the threshold nuclear recoil energy of the DAMIC silicon detector. There is no significant difference in the distribution between the brute-force ($\delta=0$) and the importance sampling results.}
	\label{ED}
\end{figure}
\begin{figure}[tbp]\centering
	\begin{subfigure}{.495\textwidth}
		\centering
		\includegraphics[width=\linewidth]{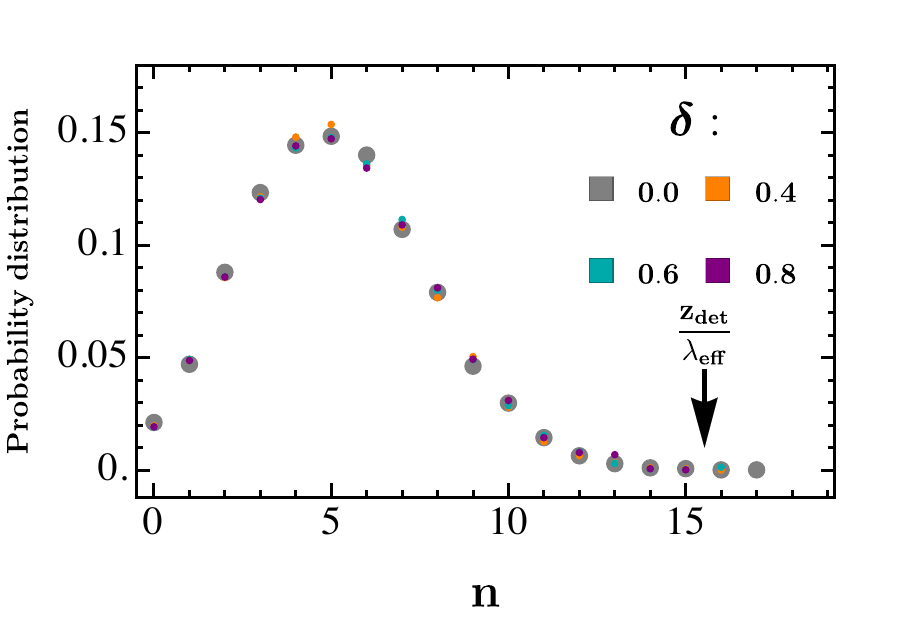}
		\caption{}
		\label{nE_D}
	\end{subfigure}\hspace{0.01cm}
	\begin{subfigure}{.495\textwidth}
		\centering
		\includegraphics[width=\linewidth]{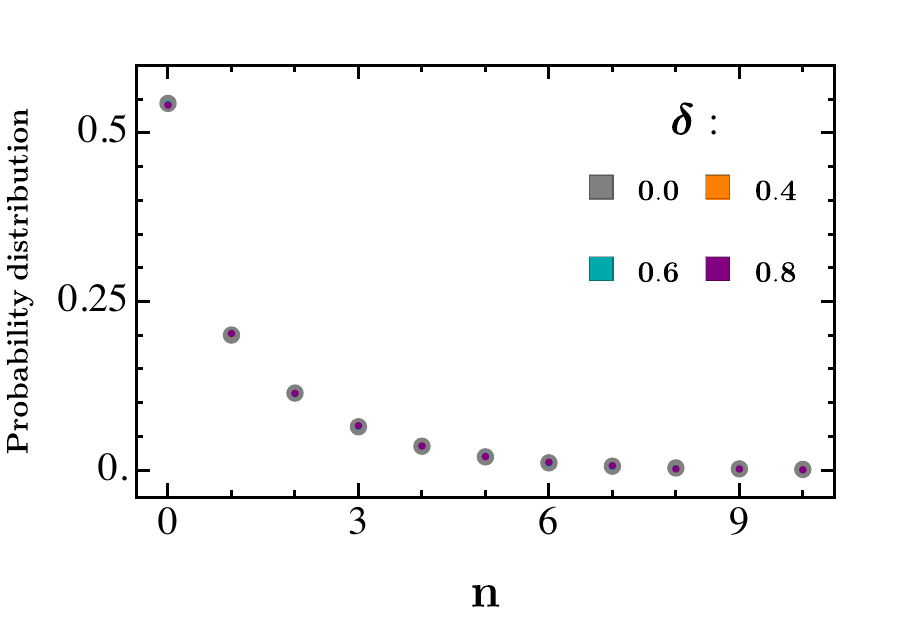}
		\caption{}
		\label{nPb_D}
	\end{subfigure}
	\caption{The distributions of the number of scatterings of capable DM particles with a mass of 1.7 GeV and a DM-proton cross section of 5.7$\times\rm{10^{-30}\,cm^2}$ while traveling through (a) the Earth's crust and (b) the lead shield. There is no significant difference in the distribution between the brute-force ($\delta=0$) and the importance sampling results.}
	\label{nD}
\end{figure}
\begin{figure}[tbp]\centering
	\begin{subfigure}{.327\textwidth}
		\centering
		\includegraphics[width=\linewidth]{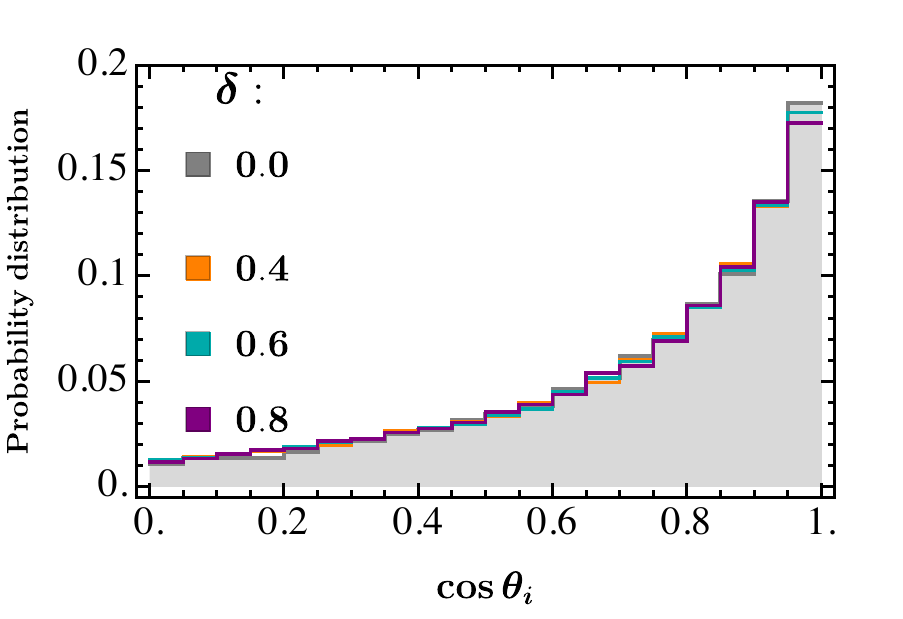}
		\caption{}
		\label{thetaE_D}
	\end{subfigure}\hspace{0.01cm}
	\begin{subfigure}{.327\textwidth}
		\centering
		\includegraphics[width=\linewidth]{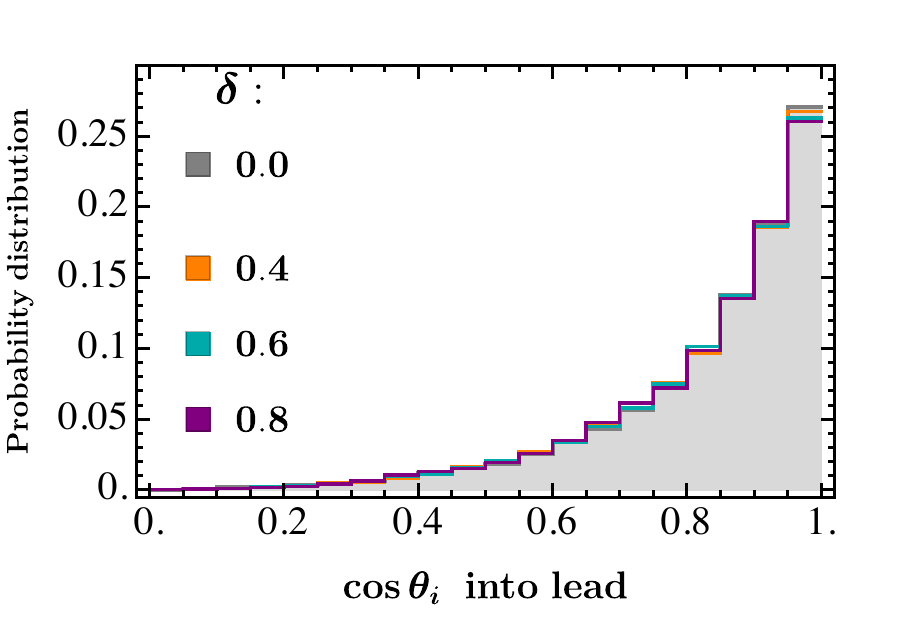}
		\caption{}
		\label{thetaPb_D}
	\end{subfigure}\hspace{0.01cm}
	\begin{subfigure}{.327\textwidth}
		\centering
		\includegraphics[width=\linewidth]{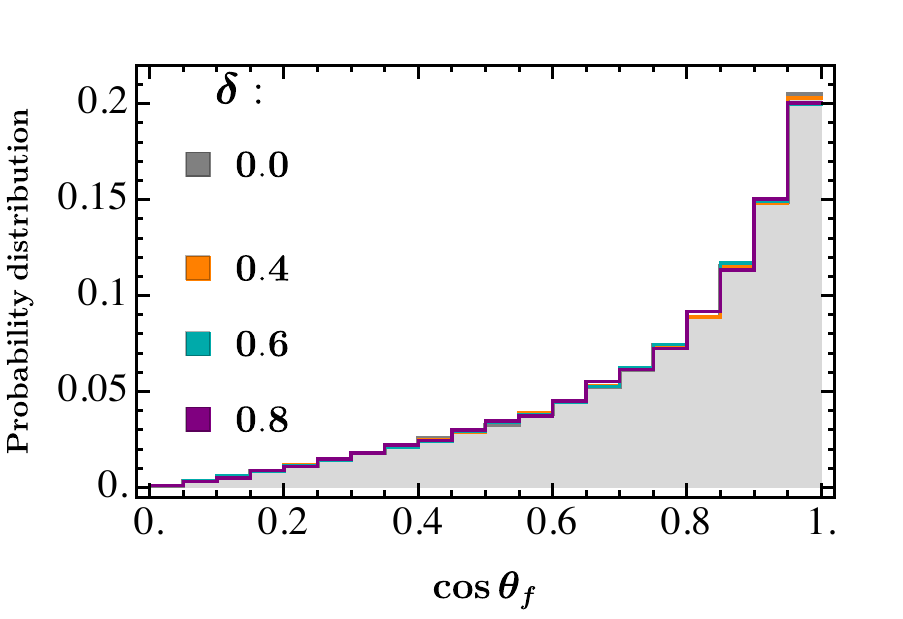}
		\caption{}
		\label{thetaSi_D}
	\end{subfigure}
	\caption{The zenith angle distributions of capable DM particles (a) on the Earth's surface, (b) on the lead shield, and (c) on the DAMIC detector with a mass of 1.7 GeV and a DM-proton cross section of 5.7$\times\rm{10^{-30}\,cm^2}$. There is no significant difference in the distribution between the brute-force ($\delta=0$) and the importance sampling results.}
	\label{theta_D}
\end{figure}

In figures~\ref{f_speedD}--\ref{theta_D}, the gray histograms show the results of the brute-force Monte-Carlo simulation using the actual step length distribution. The orange, cyan, and purple lines are the results of importance sampling Monte-Carlo simulations using the modified step length distribution with parameters $\delta=\rm{0.4, 0.6, and \,0.8}$ respectively. All the distributions are normalized to 1.
\par
Figure~\ref{vi_D} shows the initial speed distribution of capable DM particles. The initial speed distribution of all of DM particles (the light-green histogram) shows how the speed distribution of capable DM particles is significantly different than for all DM particles. The final speed distribution of capable DM particles on the surface of the DAMIC detector is shown in figure~\ref{vf_D}. The sharp cut at $v_f=$ 505 km$\cdot\rm{s^{-1}}$ corresponds to the minimum energy of a DM particle with mass 1.7 GeV to scatter off a silicon nucleus in the DAMIC detector and to deposit the threshold energy of 550 eV~\cite{bounds}. 
\par
Figure~\ref{EfEi_D} shows the distribution of the ratio of the final energy at arrival to the detector, to the initial energy at the Earth's surface of capable DM particles. We see that capable 1.7 GeV DM particles lose on average 35\% of their initial energy while passing through the Earth's crust and the lead shield.  There is no significant difference in the distribution between the brute-force ($\delta=0$) and the importance sampling results.
\par
Figure~\ref{Er_D} shows the nuclear recoil energy distribution of DM particles in DAMIC; $E_r=$ 550 eV is DAMIC's threshold nuclear recoil energy. Only 15\% of the capable DM particles which interact,  deposit energies above DAMIC's threshold.  There is no significant difference in the distribution between the brute-force ($\delta=0$) and the importance sampling results.
\par
The distributions of the number of scatterings of capable DM particles while passing through the Earth's crust and the lead shield are shown in figures~\ref{nE_D} and \ref{nPb_D} respectively. On average, capable DM particles with a mass of 1.7 GeV scatter five times in the Earth's crust and once in the lead shield. The black arrow in figure~\ref{nE_D} shows the average number of scatterings in the Earth's crust would be 15.5, just based on interaction length, if they traveled vertically and there was no biasing due to the requirement on the minimum energy at the detector.
\par
Finally, figure~\ref{theta_D} shows the zenith angle distributions of the capable DM particles on the Earth's surface, on the lead shield, and on the DAMIC detector.  One sees that large zenith angle events are less likely to reach the detector with sufficient energy, unsurprisingly.    
\par
Figures~\ref{f_speedD} -- \ref{theta_D} demonstrate significant agreement between the brute-force and the importance sampling Monte-Carlo simulations, and that our importance sampling approach reproduces all aspects of the propagation, not just the overall attenuation and final spectrum. The importance sampling results required 100 to 1000 times less computational power than the brute-force simulations depending on the magnitude of the path length modification parameter, $\delta$. As the modification parameter increases, the required computational power to reach the same statistical confidence level decreases. On the other hand, due to the presence of the hard-cut boundary conditions on the energy and the vertical displacement of the propagating DM particles, increasing the modification parameter beyond 0.6 increases the statistical noise in some quantities, notably the probability of capable trajectories with 10 or more interactions (see figure \ref{nD}).  Since these account for a small fraction of capable trajectories, a little noise here would not impair the quality of the predictions for $a_c$ and $\tilde{f}(v)$. To balance between the computational cost of the simulation and the quality of the distributions of all physical observables, we adopted the path length modification parameter of $\delta=$ 0.6 in~\cite{bounds}.

\section{Summary}
\par
In this paper we introduced and validated an importance sampling method for Monte-Carlo propagation of DM trajectories through an overburden.  With a simple modification in the sampling of just one variable, we reduced the simulation time by a factor 1000, and identified other modifications which could speed it up further if needed.   
The new method makes it computationally feasible to treat attenuation factors of $10^7$ and larger, as encountered in analyzing current experiments. This importance sampling tool will be valuable for analyzing results of the upcoming generation of experiments, for which even more severe attenuation will need to be accurately modeled.  

We explored the approximation introduced by Starkman et al.~\cite{Starkman} (SGED) to estimate the energy loss of strongly interacting particles penetrating an overburden, which was used by Kouvaris and Shoemaker~\cite{Kouvaris2014} (KS) to obtain the DM-proton cross section sensitivity limit for DAMIC.  Our detailed Monte-Carlo simulations predict $10^4$ successful events for the limiting KS cross section, when by their analysis none should be found~\cite{Kouvaris2014}.  The DM-proton cross section reach of DAMIC based on Monte-Carlo simulation of trajectories, is a factor 1.8 -- 5.6 higher than obtained by KS using the SGED approximation.   

We identified several contributing factors to the inadequacy of the SGED/KS approach.  The most important problem is that the trajectories of DM particles which arrive to the detector with sufficient energy that they are capable of triggering the detector, are highly skewed relative to average trajectories.  These trajectories have significantly fewer scatterings, longer step lengths and smaller scattering angles than average. The SGED/KS treatment of DM energy loss underestimates the length of trajectories through the overburden, overestimates the average CM scattering angle and fractional energy loss per collision of \emph{successful} DM particles, and underestimates their mean distance between interactions. The SGED\textbf{/}KS cross section estimator given in eq.~\eqref{SGEDESimple} thus involves the product of an underestimated and overestimated quantity, so the resultant cross section estimate is not a bound, contrary to the assumption of~\cite{Kouvaris2014}. 
In the Appendix, we improve the most crude version of the SGED approximation as used by KS, to use the number of events observed by the experiment rather than assuming none are observed; that makes a 15\% difference in the SGED/KS cross section estimate.   

We also examined a short-cut approach to trajectory simulation recently used by Emken, Kouvaris and Shoemaker~\cite{Emken} (EKS) for sub-GeV DM. We find a factor $\rm{4.4\times10^{6}}$ more events for 50 MeV and $\rm{\sim2.4\times10^{4}}$ more events for 1 GeV for the EKS critical cross section, than should be found if the EKS procedure were correct.
 
An interesting by-product of our validation of the importance sampling method, is the detailed characterization of the trajectories of capable and successful DM particles through the overburden. This information can guide future development of even more efficient importance sampling.

%Our analysis (section~\ref{SGED}) reveals the origin of the defect in the SGED approximation:  successful DM particles exploit the tails of the scattering-length and scattering-angle distributions to minimize their energy loss.   
%Using Monte-Carlo simulation of trajectories results in a factor 1.8 -- 5.6 stronger DM-proton cross section bound in comparison to the SGED\textbf{/}KS approximation when applied to the DAMIC experiment.  %The SGED\textbf{/}KS approximation was shown to give a poor description of the physics of the DM particles propagating in the Earth and the lead shield, by comparing attenuation factors obtained from Monte-Carlo simulation versus the SGED\textbf{/}KS approximation.  
%Several other features of successful trajectories such as number of scatterings and scattering-angle distributions are also shown to be discrepant.  %This difference corresponds to a four order-of-magnitude larger expected total number of events for a DM mass $10$ GeV at the limiting cross section. 

%As we have shown here, the inaccuracy of the SGED/KS method in this application derives from the fact that the experiments cut on the DM particle energy at a specified depth, and thus do not sample average trajectories.  Rather, a subset of trajectories  whose energy losses and path lengths are smaller than average are what matter; these make a very small contribution to the ensemble average that is described by the SGED/KS approximation. 
\par
%In section~\ref{IS} we introduced an importance sampling Monte-Carlo method to make simulation of DM propagation through an overburden computationally feasible.  In section~\ref{valid}, we demonstrate the validity of the method by comparing distributions of all relevant physical quantities simulated with the brute-force and the importance sampling methods, for a benchmark mass of 1.7 GeV. 
The DM{\scriptsize ATIS} code, which we developed for this study, will be made publicly available on-line~\cite{code} in due time.\\

\noindent {\bf Acknowledgements:}  MSM acknowledges support from the James Arthur Graduate Assistantship;  the research of GRF has been supported by NSF-PHY-1212538 and NSF-AST-1517319.

\appendix
\section{Summary of the SGED\textbf{/}KS method}\label{SGEDsum}
%If the DM-nucleon cross section is large enough, the DM particles lose too much energy in collisions prior to reaching the detector shielded by an overburden, to be able to make an energy deposit above the detector threshold. 
The first paper to point out that strongly interacting particles suffer energy loss which significantly affects their detectability was SGED~\cite{Starkman}.  They proposed an approximation to calculate the maximum cross section for which a given detector would be able to see any events at all due to this effect. KS~\cite{Kouvaris2014} applied this method for actually giving limits on the DM-proton cross section from DAMIC data. Here we provide, for reference, the SGED\textbf{/}KS calculation as applied to DAMIC. We also describe a way to improve on their max-to-min extremization, to get a more accurate limit even within the framework of their basic energy loss estimation.  We emphasize that the uncontrolled aspects of their approximations --- with the underestimated trajectory length erring in the opposite direction as the overestimated average fractional energy loss and number of interactions --- means that the method cannot be used to obtain reliable bounds or to extract cross sections if a signal is seen. Thus the Monte-Carlo approach is essential for obtaining accurate bounds. The purpose of this appendix is to quantify the impact of some inessential simplifications of the SGED\textbf{/}KS treatment.
\par
The differential energy-loss of a DM particle with mass $m$ and energy $E$ to nuclei of mass number $A$ while passing through a shielding material is
\beq
\begin{split}\label{edEA0}
	\frac{dE_A}{dz}&=-\frac{\left\langle E_{r,A}\right\rangle}{\lambda_{eff}}\;,
\end{split}\eeq
where $\lambda_{eff}\equiv(\sum_{A}n_A\sigma_A)^{-1}$ is the mean path length of DM particles in a shielding material with a mix of nuclei.  $\left\langle E_{r,A}\right\rangle$, the average nuclear recoil energy of the DM particle scattering off a nucleus of mass number $A$, is
\beq
\begin{split}\label{eErAve}
	\left\langle E_{r,A}\right\rangle = \frac{2\,\mu_A^2}{m\,m_A}\,E\;,
\end{split}\eeq
where $\mu_A$ is the DM-nucleus reduced mass.
\par
Using eqs.~\eqref{edEA0} and~\eqref{eErAve}, the total differential energy-loss of a DM particle in a scattering off a shielding material with a mix of nulcei is 
\beq
\begin{split}\label{edEdz}
	\frac{dE}{dz}&=\sum_{A}P(A)\,\frac{dE_A}{dz}\\
	&=-E\sum_{A}n_A\,\sigma_A\left( \frac{2\,\mu_A^2}{m\,m_A}\right) \;,
\end{split}\eeq
where $P(A)\equiv\frac{n_A\,\sigma_A}{\sum_A n_A\,\sigma_A}$ is the probability that the DM particle scatters off a nucleus of mass number $A$ and number density $n_A$, for DM-nucleus cross section $\sigma_A$. The spin-independent DM-nucleus cross section is related to DM-nucleon cross section, $\sigma_p$, by
\beq\begin{split}
	\label{e_sigmaAF2}
	\sigma_{A}&=\sigma_p \left(\frac{\mu_A}{\mu_p} \right) ^2A^2\,F_A^2(E_r)   \;.
\end{split}\eeq
where form factor $F(E_r)=F(q\,r_A)$ captures the effect of nuclei internal structure in DM-nucleus collisions with non-zero momentum transfer. We used analytical expression which is proposed by Helm~\cite{Helm} to calculate the nuclear form factor
\beq
\label{e_FormFactor}
F_A(E_r)=F(qr_A)=3\,\frac{\sin(qr_A)-qr_A\cos(qr_A)}{(qr_A)^3}\,e^{-(qs)^2/2}\;,
\eeq
where the effective nuclear radius $r_A$ can be approximately found by fitting the muon scattering data to a Fermi distribution~\cite{fricke1995}
\beq
\label{e_NuclearRadius}
r_A^2=c^2+\frac{7}{3}\pi^2a^2-5s^2\;,
\eeq
with parameters: $c\simeq(1.23A^{1/3}-0.6)$ fm, $a\simeq0.52$ fm, and $s=0.9$ fm. 
\par
Neglecting the momentum transfer dependence of the DM-nucleus cross section, i.e., $F_A^2(E_r)=1$, the energy of a DM particle after traveling a distance $L$ in the continuous energy-loss approximation is
\beq
\label{eEL}
E(L)=E(0)\,\exp\left( -\frac{2\,\sigma_p}{m}(\rho_S\mathscr{F}_{S}L)\right) \;,
\eeq
where $E(0)$ is the initial energy of the DM particle before entering the shielding material of mass density $\rho_S$ and $\mathscr{F}_S=\sum_A f_A\left( \frac{\mu_A^2}{m_p\,\mu_p}\right)^2$.
\par
SGED\textbf{/}KS assumes the path length can be replaced by the depth of the detector. So, the final energy of a DM particle for DAMIC detector, operating at an underground depth of $z_{det}=$ 106.7 meters and shielded by a lead shield of thickness $z_{Pb}=$ 6 inches, is
\beq
\label{eSGEDE}
E(z_{det},z_{Pb})=E(0)\,\exp\left( -\frac{2\,\sigma_p}{m}(\rho_E\mathscr{F}_{E}z_{det}+\rho_{Pb}\mathscr{F}_{Pb}z_{Pb})\right) \;,
\eeq
where q.~\eqref{eSGEDE} is a detailed version of eq.~\eqref{SGEDESimple} accounting for multiple components in the overburden.
\par
The second simplification they use is to find the DM-proton cross section for which the DM particle with the biggest possible energy on the Earth surface, $E(0)=E_{max}=\frac{1}{2}m\,v^2_{max}$, is able to potentially trigger the detector underground with the recoil energy threshold of the detector. DAMIC's maximum cross section using this method, which totally ignores the observed number of events, is
\beq
\label{eSGEDb}
\sigma_p^{max}=\frac{m}{2\,(\rho_E\mathscr{F}_{E}z_{det}+\rho_{Pb}\mathscr{F}_{Pb}z_{Pb})}\ln \left(\frac{E_{max}}{E_{min}}\right) \;,
\eeq
where $E_{min}$ is the minimum energy of a DM particle to scatter a silicon target nucleus in the DAMIC detector and produce the threshold recoil energy $E_r^{th}$
\beq
\label{eEmin}
E_{min}=\frac{(m+m_{Si})^2}{4\,m \, m_{Si}} E_r^{th}\;.
\eeq
\par
An improvement on the SGED\textbf{/}KS method is to find the cross section which predicts the observed number of events in the detector. This can be done as follows. As we showed in~\cite{bounds}, the expected total number of events for DAMIC's silicon target ($A=28$) is
\beq\begin{split}
	\label{e2_NdEr}
	N(\sigma_p,\,z_{det},\,z_{Pb})=&\,t_e\,M_A\,A^2\,\frac{\rho\,\sigma_p}{4\,m\,\mu^2_p}\int_{0.55\,keV}^{7\,keV}F_A^2(E_r)\,dE_r\int_{v'_{min}(E_r,A)} \frac{f'(\vec{v}',\vec{v}_{det})}{v'}\,d^3v'\;,
\end{split}\eeq
where $f'(\vec{v}',\vec{v}_{det})$ is the normalized velocity distribution of DM particles on the surface of the detector. The lower limit of integration, $v'_{min}$, is the minimum speed that a DM particle on the detector needs in order to deposit the recoil energy $E_{r}$ into the detector:
\beq
\label{e_v'm}
v'_{min}(E_r,A)=\sqrt{\frac{m_A\,E_{r}}{2\,\mu^2_{A}}}\;.
\eeq
\par
For DM mass $m$ and trial DM-proton cross section $\sigma_p$, velocities of DM particles on the Earth's surface, $v$, scale by a common factor to $v'=k(\sigma_p)\,v$ on the detector surface where
\beq
\label{eSGEDk}
k(\sigma_p)=\exp\left( -\frac{\sigma_p}{m}(\rho_E\mathscr{F}_{E}z_{det}+\rho_{Pb}\mathscr{F}_{Pb}z_{Pb})\right) \;,
\eeq
where this scale factor is normalized to unity at zero DM-nucleon cross section, i.e. $k(0)=1$.
Using eqs.~\eqref{e2_NdEr} and~\eqref{eSGEDk}, the expected total number of events for DAMIC is
\beq\begin{split}
	\label{e_NdEr}
	N(\sigma_p,\,z_{det},\,z_{Pb})=&\,t_e\,M_A\,A^2\,\frac{\rho\,\sigma_p}{4\,m\,\mu^2_p\,k(\sigma_p)}\int_{0.55\,keV}^{7\,keV}F_A^2(E_r)\,dE_r\int_{v_{min}(E_r,\,A,\,\sigma_p)} \frac{f(\vec{v},\vec{v}_{det})}{v}\,d^3v\;,
\end{split}\eeq
where $z_{det}$ and $z_{Pb}$ dependences are implicit in definitions of $k(\sigma_p)$ and $v_{min}(E_r,\,A,\,\sigma_p)$. $f(\vec{v},\vec{v}_{det})$ is the normalized velocity distribution of DM particles on the Earth's surface. The lower limit of integration, $v_{min}$, is the minimum speed that a DM particle on the Earth's surface needs in order to deposit the recoil energy $E_{r}$ into the detector:
\beq
\label{e_vm}
v_{min}(E_r,\,A,\,\sigma_p)=k^{-1}(\sigma_p)\sqrt{\frac{m_A\,E_{r}}{2\,\mu^2_{A}}}\;. 
\eeq
\par
\begin{figure}[tbp]\centering
	\includegraphics[width=0.8\linewidth]{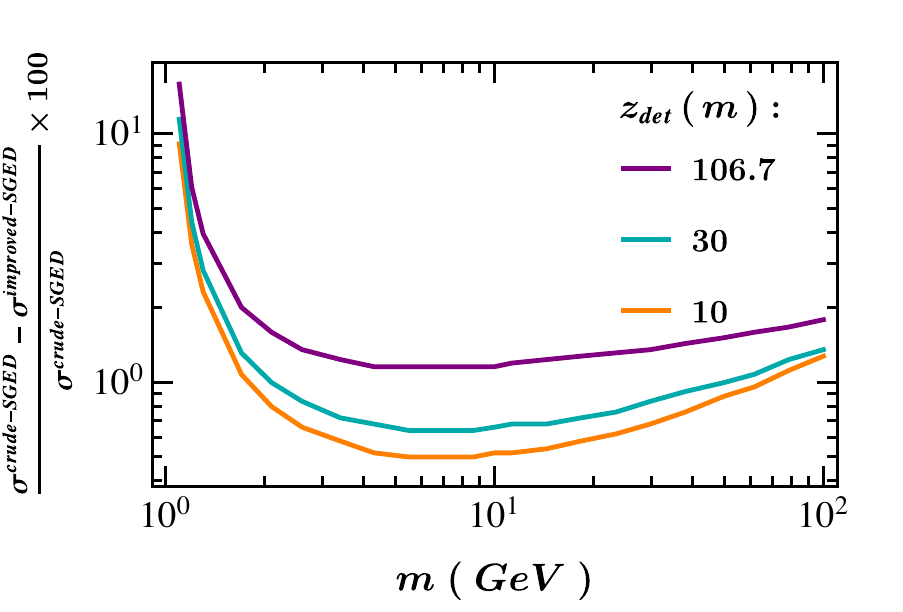}
	\caption{\label{SGED_rel_deviation} The percentage change of the DAMIC bounds using the crude SGED\textbf{/}KS method in comparison to the improved SGED\textbf{/}KS method of the DAMIC bounds if the collected at three different detector depths: 10, 30, 106.7 meters underground.}
\end{figure}
As the trial cross section is decreased from some large value, the expected total number of events, calculated using eq.~\eqref{e_NdEr}, is increasing monotonically. The improved SGED\textbf{/}KS 90\% lower bound on the allowed (by DAMIC) cross section would then be defined to be the value of $\sigma_p$ for which the expected total number of events is 123 (The equivalent 90\% CL upper limit on 106 observed total number of events is 123). Figure~\ref{SGED_rel_deviation} shows the percentage change of the DAMIC bounds using the crude SGED\textbf{/}KS method in comparison to the improved SGED\textbf{/}KS method if the collected at three different detector depths: 10, 30, 106.7 meters underground. This shows that the crude SGED\textbf{/}KS method overestimates the maximum excluded cross section by up to $15\%$ in comparison to the improved SGED\textbf{/}KS method.
\par
The attenuation parameter of the total number of events in the improved SGED\textbf{/}KS method, which is shown with an orange solid line in figure~\ref{DAMIC_att@sigma_SGED}, is
\beq
\label{aN_SGED}
a_N = \frac{N(\sigma_p,\,z_{det},\,z_{Pb})}{N(\sigma_p,\,0,\,0)}\;.
\eeq
\par
The fraction of capable DM particles at the detector's surface, $\eta(\sigma_p,\,z_{det},\,z_{Pb})$, is
\beq
\label{eta_SGED}
\eta(\sigma_p,\,z_{det},\,z_{Pb}) = \int_{v_{min}(\sigma_p,\,E_r,\,A)} f(\vec{v},\vec{v}_{det})\,d^3v\;,
\eeq
where mass, $z_{det}$ and $z_{Pb}$ dependences are implicit in definition of $v_{min}$ in eq.~\eqref{e_vm}. $E_r=550$ eV and $A=28$ should be used for DAMIC. 
\par
The attenuation parameter of capable particles $a_c$ in the improved SGED\textbf{/}KS method, which is shown with an orange dashed line in figure~\ref{DAMIC_att@sigma_SGED}, is
\beq
\label{ac_SGED}
a_c = \frac{\eta(\sigma_p,\,z_{det},\,z_{Pb})}{\eta(\sigma_p,\,0,\,0)}\;.
\eeq
\bibliography{g1}
\end{document}